\documentclass[10pt,letterpaper]{article}
\usepackage{opex3}
\usepackage{graphicx}
\usepackage{cite}

\begin{document}

\title{Imaging using quantum noise properties of light}

\author{Jeremy B. Clark,$^{1*}$ Zhifan Zhou,$^{1,2}$ Quentin Glorieux,$^{1}$
\newline Alberto M. Marino,$^{1}$ and Paul D. Lett$^{1}$}

\address{$^{1}$Quantum Measurement Division,~National Institute of Standards and Technology,
\\and Joint Quantum Institute, NIST and University of Maryland, 100 Bureau Dr., Gaithersburg, MD 20899-8424, USA

$^{2}$Quantum Institute for Light and Atoms, State Key Laboratory of Precision Spectroscopy, Department of Physics, East China Normal University, Shanghai 200062, China}

\email{$^{*}$jeremy.clark@nist.gov} 



\begin{abstract}
We show that it is possible to estimate the shape of an object by measuring only the fluctuations of a probing field, allowing us to expose the object to a minimal light intensity.
This scheme, based on noise measurements through homodyne detection, is useful in the regime where the number of photons is low enough that direct detection with a photodiode is difficult but high enough such that photon counting is not an option.
We generate a few-photon state of multi-spatial-mode vacuum-squeezed twin beams using four-wave mixing and direct one of these twin fields through a binary intensity mask whose shape is to be imaged.
Exploiting either the classical fluctuations in a single beam or quantum correlations between the twin beams, we demonstrate that under some conditions quantum correlations can provide an enhancement in sensitivity when estimating the shape of the object.
\end{abstract}




\section{Introduction}
The use of quantum correlated resources to enhance sensing and measurement has been an active field of research in recent years \cite{giovannetti_advances_2011}.
Multi-spatial-mode quantum states of light have become available \cite{kolobov_spatial_1999,kolobov_quantum_2006,corzo_multi-spatial-mode_2011,boyer_generation_2008} and have allowed for the development of imaging techniques that can surpass classical limits on resolution \cite{kolobov_quantum_2000,treps_quantum_2003,giovannetti_sub-rayleigh-diffraction-bound_2009} or enhance signal-to-noise \cite{treps_surpassing_2002,levenson_reduction_1993,brambilla_high-sensitivity_2008}.
Here we present a technique that exploits the noise properties of multi-spatial-mode vacuum-squeezed twin beams to detect the shape of an object being probed by one of the beams.
While we are able to estimate the shape of the object by monitoring only the fluctuations of the beam that probes it, we obtain an enhancement in this estimation sensitivity under certain conditions by exploiting the quantum correlations between the twin beams.

It is possible to estimate the shape of a binary (hard-apertured) intensity mask by probing it with a light field and performing a homodyne detection of the light that passes through it.
The local oscillator (LO) spatially selects which portions of the probing field are detected, so the homodyne detection effectively acts as an imaging device \cite{lugiato_spatial_1993,gatti_quantum_1995}.
By optimizing the shape of the LO to maximize the mode matching with the probing field after the mask (the LO-mask overlap), the object's shape can be inferred.
If the mask is probed with a state of light consisting of a small number of photons, determining the mean value of the field becomes difficult due to unavoidable quantum noise.
We can, however, use the same homodyne technique to measure the noise of the field rather than its mean value in order to extract information about the object's shape.
Homodyne detection provides the additional advantage of allowing the discrimination of a signal against a background at other wavelengths.

A coherent state would not be a suitable choice to probe the mask since it is a displaced vacuum state.
Its noise properties would thus remain unchanged regardless of how the probe field is altered or attenuated by the mask \cite{marino_extracting_2012}.
On the other hand, a state whose quadrature fluctuations are not at the shot noise limit (SNL) can provide information about the shape of the mask.
For example, if a thermal state is used, the mask's shape can be estimated by manipulating the shape of the LO to optimize the geometric overlap between the field that passes through the mask and the LO to maximize the detected level of excess noise relative to the SNL.
Since thermal states are classically obtainable, we will refer to such a single-beam method as a classical noise imaging technique.

In many situations a reference beam is used to eliminate correlated sources of noise and to improve the signal-to-noise ratio for a given measurement \cite{brida_giorgio_experimental_2010, abouraddy_role_2001}.
We extend this idea to noise imaging by exploiting the quantum correlations between vacuum-squeezed twin beams to improve the estimation of the shape of the mask over the classical noise imaging technique described above.
We refer to this two-beam technique as a quantum noise imaging technique.
In this paper, we use twin beams generated by four-wave mixing (4WM) to directly compare these two techniques and find that the quantum technique provides an enhanced sensitivity to changes in the LO-mask overlap and thus a greater confidence in estimating the shape of the mask.

\section{Four-wave mixing and squeezed light detection}
\begin{figure}[htbp]
\centering\includegraphics[width=\columnwidth]{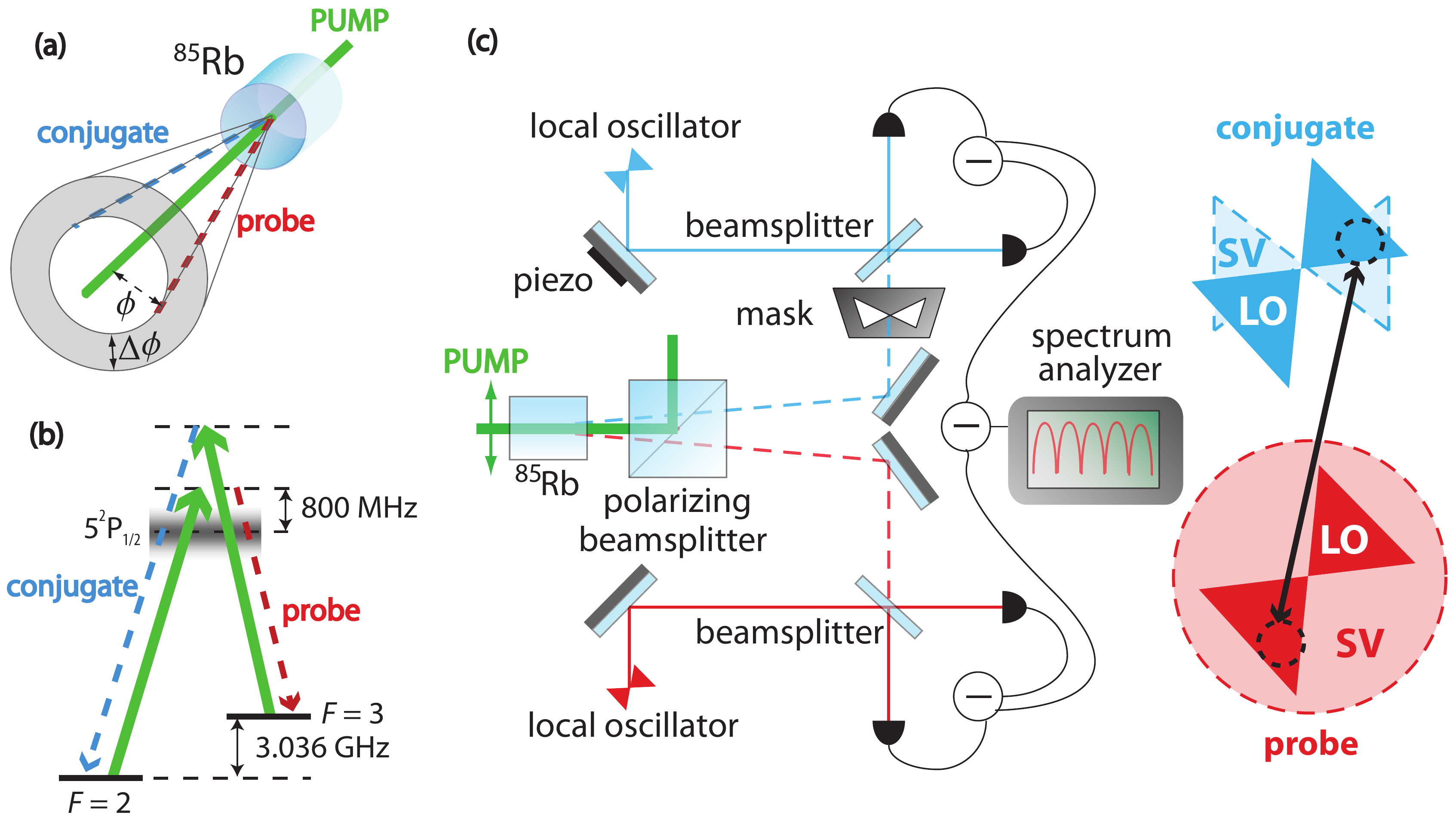}
\caption{\label{fig:4WMandsetup}(a) Four-wave mixing in $^{85}$Rb. Probe and conjugate fields are coupled over a range of angles $\Delta \phi$ and selected for measurement via homodyne detection at $\phi$. (b) \nobreakdashes Double\nobreakdashes-$\Lambda$ scheme in $^{85}$Rb. (c) Experimental set-up.  Probe and conjugate local oscillators are shaped into bow ties and are rotated with respect to the mask. Resultant homodyne signals are subtracted.  The black dotted circles joined by the arrow indicate coherence areas, localized regions of correlations between the probe and conjugate.  LO denotes the local oscillator and SV the squeezed vacuum light.}
\end{figure}

To generate the light states needed for this experiment, we have used 4WM in a double-$\Lambda$ configuration \cite{glorieux_double-_2010} in a hot $^{85}$Rb vapor (Fig.~\ref{fig:4WMandsetup}(a)-(b)).
The 4WM process converts two photons from a single pump beam into a pair of photons emitted into twin fields referred to as the probe and conjugate.
These twin beams exhibit strong amplitude correlations such that the amplitude difference noise is below the SNL \cite{mccormick_strong_2008}.
Taken individually, these fields are thermal states exhibiting uniformly distributed excess noise.

Quantum correlations between the probe and the conjugate can be measured by performing a balanced homodyne detection of each field and subtracting the homodyne signals for appropriately chosen phases of the LOs.
The noise of a single mode's generalized quadrature $\hat{X}_{\theta}=\frac{1}{\sqrt{2}}(\hat{a}^{\dagger}e^{i\theta}+\hat{a}e^{-i\theta})=\hat{X}\cos\theta+\hat{Y}\sin\theta$ is measured by choosing the LO phase $\theta$.
We subtract the noise of the probe and conjugate generalized quadratures to obtain the noise of the joint quadrature $
\hat{X}^{J}_{\theta}=\hat{X}^{p}_{\theta_p}-\hat{X}^{c}_{\theta_c}$.
Given probe and conjugate LO phases respectively denoted by $\theta_{p}$ and $\theta_{c}$, $\hat{X}^{J}_{\theta}$ is squeezed under the condition that $\theta_{p} + \theta_{c} = \pi$.
We therefore only need to control the phase of one of the LOs (using a PZT-mounted mirror) to detect the level of squeezing between the probe and conjugate.

To implement the 4WM process, a linearly polarized intense pump beam (300 mW) is focused down to a $\frac{1}{e^{2}}$ diameter of 1.2 mm inside a 1.25 cm long $^{85}$Rb cell heated to 110\ensuremath{^\circ}C.
The probe and conjugate frequencies are unseeded, so the resulting probe and conjugate fields are generated from spontaneous emission.
The frequency of the pump beam is detuned 800 MHz to the blue of the $\vert5^{2}S_{1/2},F~=~2\rangle \rightarrow \vert5^{2}P_{1/2},F~=~3\rangle$ transition at 795~nm.
After separating the pump beam from the probe and conjugate beams with a polarizing beam splitter, each beam is sent to a balanced homodyne detector using a pair of matched photodiodes with quantum efficiencies of approximately 95\%.  The total optical path losses from the 4WM process to the homodyne detection are approximately 4\%.


\section{Experimental procedure}
To characterize each technique's sensitivity we reduced the LO search to a one-dimensional problem by choosing our mask to be in the shape of a bow tie that can be oriented at an arbitrary angle around its center of symmetry (Fig.~\ref{fig:4WMandsetup}(c)).
The position of the center of symmetry was fixed, and the only free parameter to be estimated was the mask's angular orientation. 
We generated probe and conjugate LOs with congruent bow tie shapes but rotated at an arbitrary common angle relative to the mask.
The estimated parameter is the overlap between the conjugate after the mask and its LO, which is a function of the angle.
We then recorded the resultant quadrature noise powers with a spectrum analyzer as a function of the overlap for the classical and quantum noise imaging techniques.
In order to reliably measure the amount of squeezing in the quantum case, the phases of the LOs were quantum noise locked  \cite{mckenzie_quantum_2005} such that the twin beam quadrature noise difference was always minimized for each overlap value.

\begin{figure}[htbp]
\centering\includegraphics[width=.75\columnwidth]{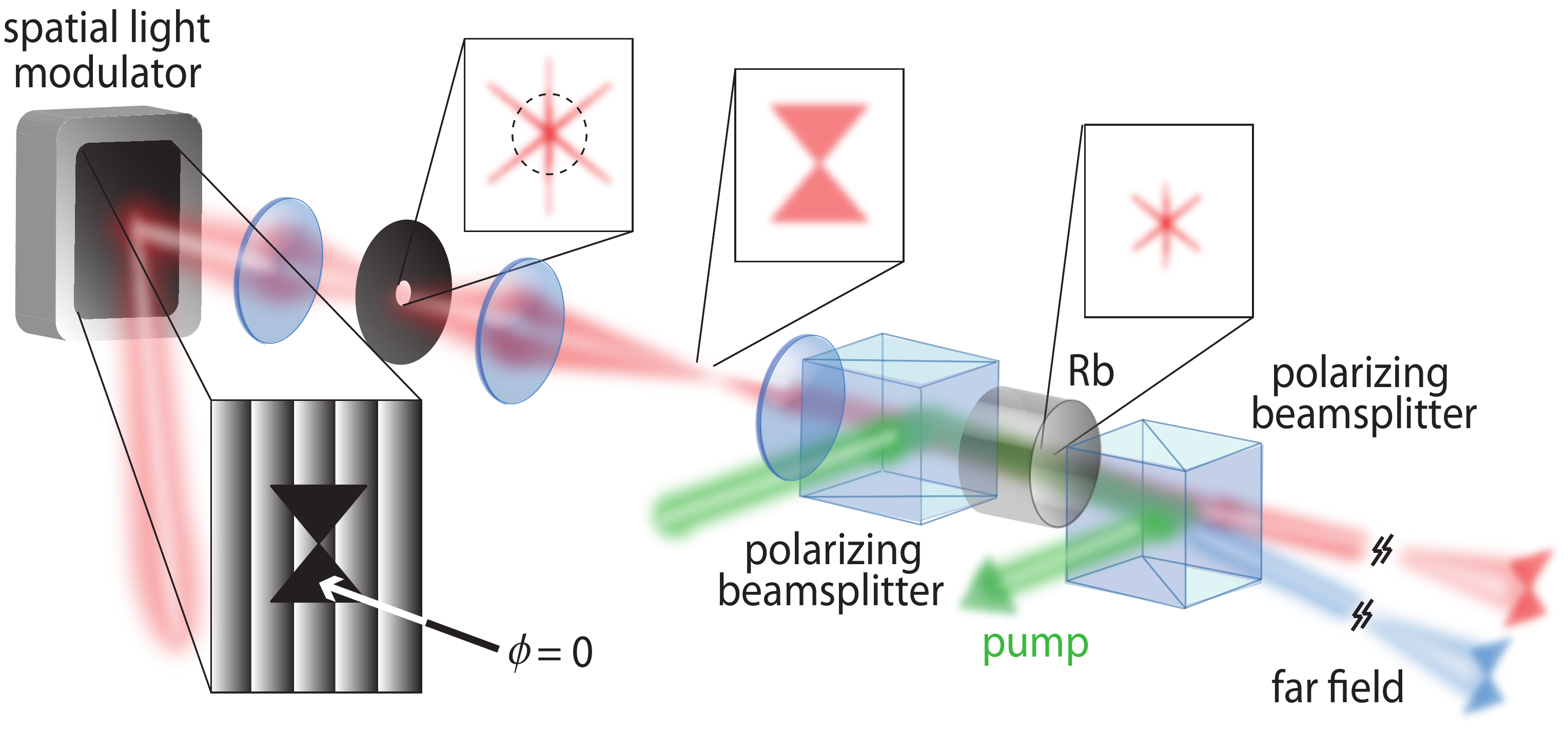}
\caption{\label{fig:SLM} A spatial light modulator writes a diffraction grating with a region of constant phase in the desired transverse shape of the local oscillator.
A coherent state is scattered from the diffraction grating and an f-f optical system and telescope focus the Fourier transform of the pattern into a Rb cell where 4WM occurs, generating two local oscillators at the appropriate frequencies and desired shape in the far field.}
\end{figure}

An important feature of the correlated fields generated by our 4WM set-up is that they are multi-spatial-mode \cite{boyer_entangled_2008}, which can in general provide greater flexibility in obtaining an enhancement with these states \cite{marino_extracting_2012}.
Quantum correlations between the probe and conjugate are localized to pairwise correlated regions within the beams referred to as coherence areas \cite{martinelli_experimental_2003,gatti_correlated_2004} (Fig.~\ref{fig:4WMandsetup}(c)).
In order to measure and compare only correlated coherence areas we symmetrically generate LOs for the probe and conjugate by shaping the probe beam using a spatial light modulator (SLM) to shape the spatial profile of a coherent state at the probe frequency.
We then use this beam to seed a second 4WM process (Fig.~\ref{fig:SLM}).
This technique \cite{kim_quadrature-squeezed_1994} ensures that only the spatial modes supported by the process will be present in the LOs and that any mode distortions due to Kerr lensing \cite{boyer_entangled_2008} will be accounted for as the beams propagate to the far field.
Additionally, the use of a second 4WM process guarantees that the spatial profiles of the probe and conjugate LOs are manipulated synchronously so that only correlated regions of the probe and conjugate fields are detected as the shapes of the LOs are manipulated.
Without any mask present, we align the LOs to the twin fields in order to maximize the squeezing.
The mask is then inserted into the conjugate field, and we begin the search for the mask's shape by manipulating the transverse profile of the LOs.

We calibrated the LO-mask overlap by splitting the seed beam used to generate the LOs and sending it to seed the 4WM process formerly used to generate the vacuum-squeezed twin beams, so that bright twin beams are obtained in their place.
These bright twin beams can then be interfered with their LOs in order to optimize mode matching and ensure proper initial alignment for the homodyne detection.
As we rotated the angular orientation of the bow tie shapes on the SLM, we observed that the homodyne mode-matching efficiencies between the bright beams and their LOs stayed near 97\% for both the probe and conjugate.
We then inserted the mask into the bright seeded conjugate beam and directly detected the fraction of transmitted power as the angle of the bow tie was changed using the SLM.
This procedure allowed us to factor any non-uniformity in the beam's intensity profile into the evaluation of the overlap at any LO orientation.

The noise generated by the homodyne difference signal was detected at 750 kHz using a spectrum analyzer in zero-span mode with a resolution bandwidth of 30 kHz, a video bandwidth of 100 Hz, and a sweep time of 1~s.
The detected noise was digitized into 460 points, and the average noise power of these 460 points was taken to be a single quadrature noise measurement.
To characterize each measurement's statistical uncertainty, we divided each trace into 46 segments consisting of 10 points each.
We adopted this procedure after verifying that clustering the data into 10 point segments maintained statistical independence among tthese segments.
The average value of each segment was tabulated, and the standard deviation of these 46 values was calculated to characterize the measurement's statistical uncertainty.
Finally, 10 series of measurements were taken at 15 different local oscillator bow tie angles for both the quantum and classical noise imaging techniques.
This data is presented for the classical and quantum techniques in Fig.~\ref{fig:sensitivities}(a).
Since each series of measurements was separated by approximately 20 minutes, the data in Fig.~\ref{fig:sensitivities}(a) indicates an experimental stability over several hours.

We first consider the quantum noise imaging technique (blue triangles in Fig.~\ref{fig:sensitivities}).
If the conjugate LO does not spatially overlap with the conjugate field that passes through the mask its measured fluctuations will be near the SNL.
On the other hand, since the probe field passes unobstructed to its homodyne detector, its homodyne detection will produce excess noise.
Therefore the homodyne difference signal will yield a net excess noise relative to the twin beam SNL.
As the overlap between the conjugate and its LO increases, however, the quantum correlations between the twin beams are recovered.
Assuming that the noise properties are identical for all the coherence areas, the search for the mask's shape is complete when the amount of squeezing is maximized for the largest LO area possible.

The classical noise imaging technique is implemented by simply blocking the probe beam (red squares in Fig.~\ref{fig:sensitivities}).
The only difference is that the conjugate beam, taken alone, is a thermal state with excess noise.
In this case, the search for the shape of the mask consists of maximizing the excess noise by manipulating the LO.

\section{Results}
\begin{figure}[htbp]
\centering\includegraphics[width=\columnwidth]{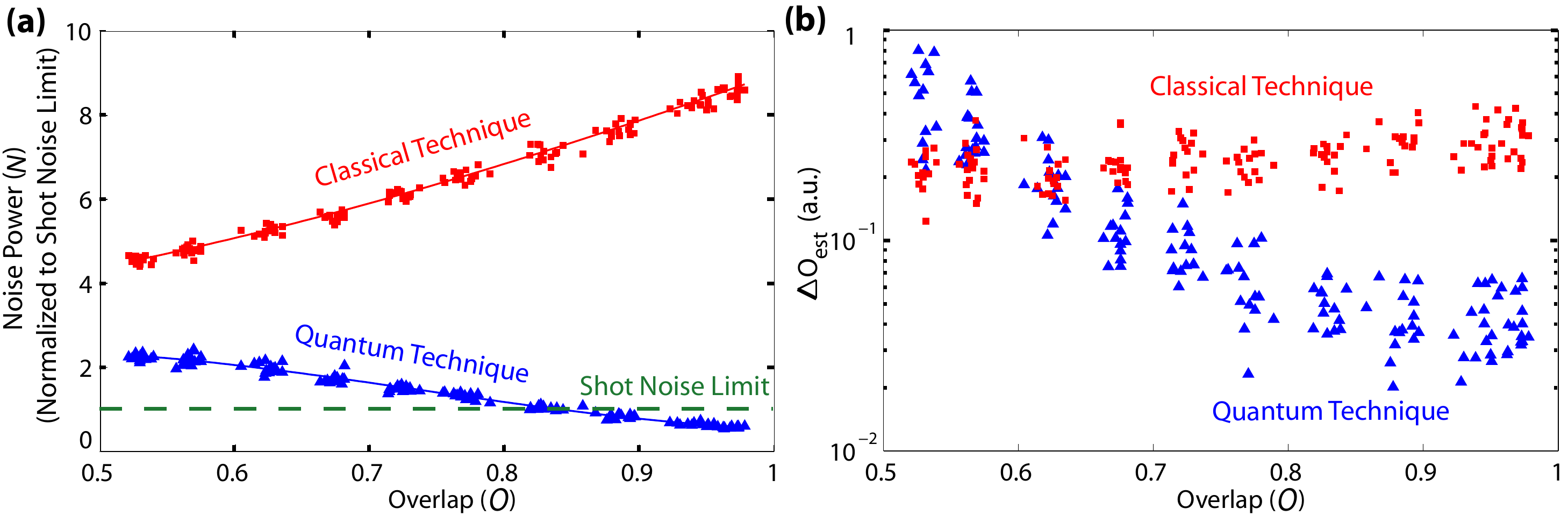}
\caption{\label{fig:sensitivities}(a) Mean quadrature noise power $N$ for the conjugate's excess noise (red squares) and twin beam difference signal (blue triangles) as a function of overlap between the mask and the conjugate LO.  Standard deviations of the mean are on the order of the marker size and not illustrated. Third order polynomial fits to the data are included. (b) Comparison of the uncertainty in the estimated LO-mask overlap for the classical and quantum techniques, on a log scale, as a function of the overlap for the single and twin beam measurements.}
\end{figure}

We wish to quantify the relative sensitivity between the quantum and classical noise imaging techniques in detecting changes in the LO-mask overlap for different overlaps.
To do this we introduce the relative uncertainty in the overlap estimation $\Delta O_{est}$:
\begin{equation}
\label{eq:errorprop}
\Delta O_{est}=\frac{\Delta N}{|\frac{\partial N}{\partial O}|}.
\end{equation}
$\Delta N$ represents the measured standard deviation of a given noise power $N$ and $\frac{\partial N}{\partial O}$ is the slope of the noise power as a function of the overlap. 
In other words, $\Delta N$ quantifies the ``noise on the noise," which incorporates both sources of statistical uncertainty and technical noise.
$\frac{\partial N}{\partial O}$ is set by both the spatial mode composition of the light illuminating the mask as well as the mask's shape \cite{marino_extracting_2012}.
To estimate $\frac{\partial N}{\partial O}$, we computed a linear fit to the noise data at overlaps greater than 0.8 and extrapolated to an overlap of unity. We then included this extrapolated point in a 3rd order polynomial fit and plotted the result in Fig. 3(a).
Fig.~\ref{fig:sensitivities}(a) confirms that for overlaps near unity the excess noise and squeezing are maximized for the classical and quantum techniques, respectively.

In Fig.~\ref{fig:sensitivities}(b), we present curves comparing $\Delta O_{est}$ for the classical and quantum schemes.
As Fig.~\ref{fig:sensitivities}(b) suggests, the quantum noise imaging technique provides a higher sensitivity to changes in the overlap than the classical technique.
Although the comparable magnitudes of $|\frac{\partial N}{\partial O}|$ indicated in Fig.~\ref{fig:sensitivities}(a) might suggest that the quantum and classical cases should exhibit comparable values for $\Delta O_{est}$, the difference between the curves in Fig.~\ref{fig:sensitivities}(b) can be explained by the variation of $\Delta N$ with overlap.
Specifically, the value of $\Delta N$ is expected to scale with the noise power $N$.
For a given overlap the magnitudes of $|\frac{\partial N}{\partial O}|$ are similar for each technique,  but the smaller value of $\Delta N$ for the quantum technique will yield a smaller value of $\Delta O_{est}$.

Since the objective of these techniques is to estimate the shape of the mask, we wish to operate in the regime where the overlap is close to unity.
For overlaps of 0.9 and above, the quantum case provides an enhancement over the classical case by a factor of 6.3 $\pm$ 0.4, corresponding to enhancement in the estimation of the angle by a factor of 3.8 for small angles.
The enhancement factor of 6.3 was calculated by averaging $\Delta O_{est}$ for overlaps above 0.9 in Fig.~\ref{fig:sensitivities}(b) for the quantum and classical techniques.
The angular enhancement factor of 3.8 was obtained by separately calibrating the measured LO-mask overlap as a function of the angle between the LO bow tie and the mask.
The indicated uncertainties here and in the figures represent one standard deviation, combined statistical and systematic uncertainties.

We have verified that the classical technique is limited by fluctuations of the conjugate field.
On the other hand, we are able to observe excess technical noise introduced by the quantum noise lock used for the measurement of the squeezed noise power for the quantum technique.
This leads to a degradation in sensitivity when estimating the angle of the mask.
Since this technical noise is not present on the excess noise measurements of the classical technique it decreases the enhancement provided by quantum noise imaging technique over the classical technique.
Accordingly, even larger enhancements should be possible.
The advantage that can be obtained with the quantum technique over the single-beam classical technique is dependent on the amount of squeezing present\cite{marino_extracting_2012}.
Losses after the twin beam source will erode this advantage depending on how they are distributed between the probe and conjugate.
In particular, unbalanced losses between the beams would alter the sensitivity and could eliminate any advantage.

\section{``Alphabet gun" test}

\begin{figure}[htbp]
\centering\includegraphics[width=.5\columnwidth]{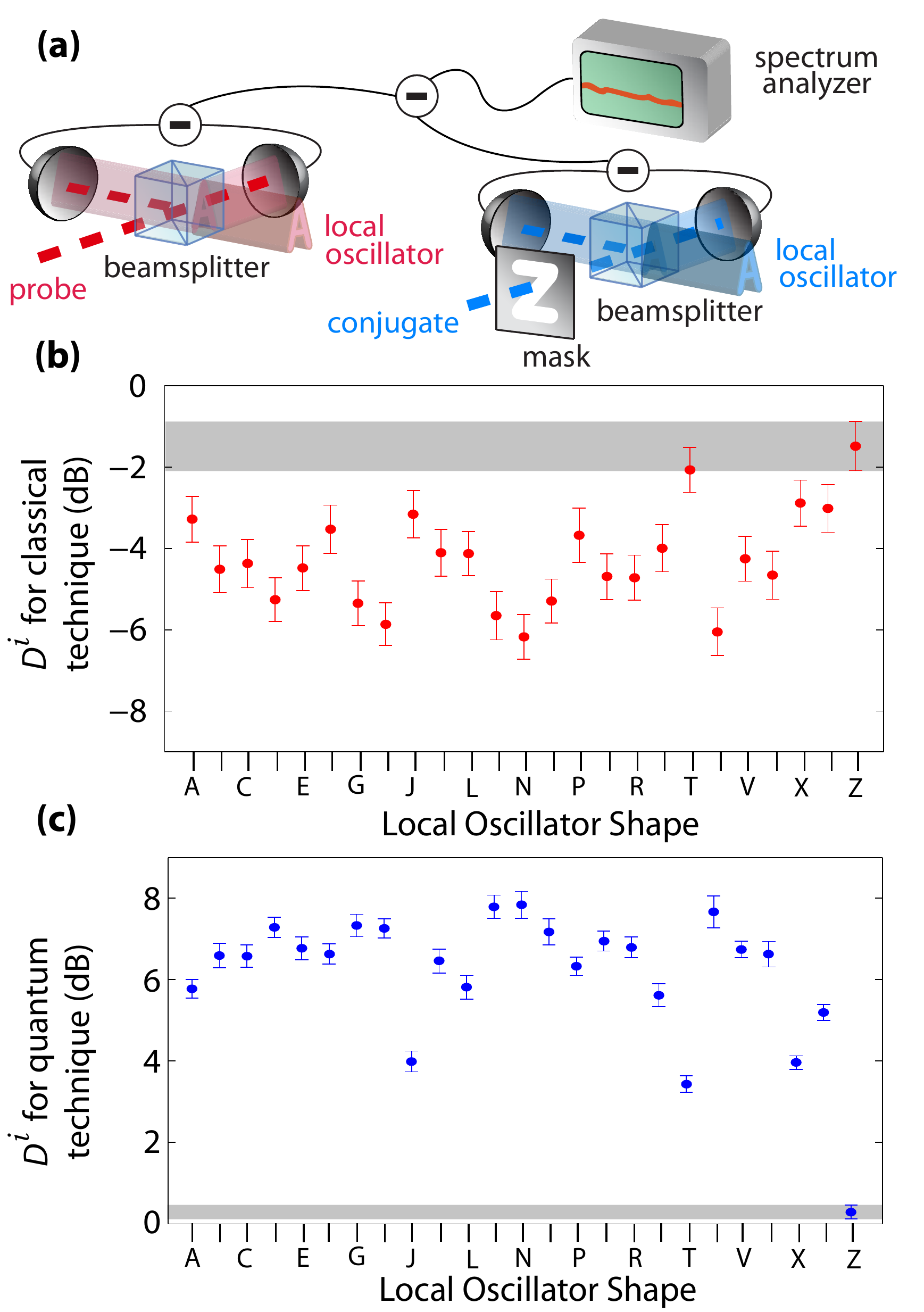}
\caption{\label{fig:zmaskdata}(a) Probe and conjugate LOs are generated in the shape of each letter of the alphabet and used in a balanced homodyne detection.  The conjugate field illuminates the mask whose shape is to be estimated. (b) Deviation from initial excess noise for the classical noise imaging technique upon insertion of the mask versus choice of letter for the local oscillator.  (c) Deviation from initial squeezing level for the quantum noise imaging technique upon insertion of the mask. The baseline of squeezing between the twin beams for the LO letter Z is -2.2~dB, measured without the mask.  With the mask inserted, only the mask shaped as the letter Z maintains any squeezing.  The gray regions in (b)-(c) represent the value of $D^{i}$ and its associated uncertainty for the correct LO choice, Z.  The letter ``I" in the chosen font did not produce a bright enough local oscillator to elevate the quadrature noise power above the electronic noise floor of our detectors.}
\end{figure}

\begin{figure}[htbp]
\centering\includegraphics[width=.25\columnwidth]{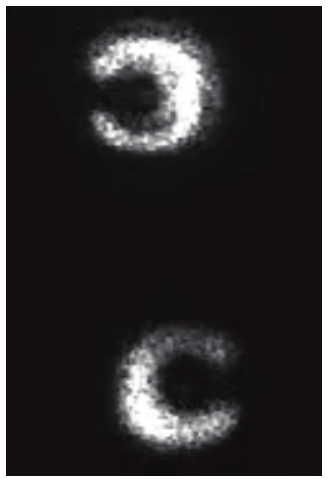}
\caption{\label{fig:LOs}Local oscillators used for the alphabet gun test.}
\end{figure}

While we have shown that the quantum noise imaging technique can improve the sensitivity in determining the maximum LO-mask overlap, we also show that this technique can be used with a simple search algorithm to recognize a mask among a given set of choices.
We place a mask cut into the shape of the letter Z into the path of the conjugate field and then shape the probe and conjugate LOs with the SLM to form the various letters of the Latin alphabet (Fig.~\ref{fig:zmaskdata}(a)).
All letters are positioned such that their glyph widths exactly overlap (Fig.~\ref{fig:LOs}).
Each letter essentially functions as a guess for the shape of the mask.
With no mask inserted into the path of the conjugate beam, each choice of LO letter yields slightly different levels of detected single beam excess noise and twin beam squeezing due to the spatial mode structure of the twin beams.
These noise levels serve as baselines for comparison when the mask is inserted.
We then look at the deviation from these baselines upon insertion of the mask:
\begin{equation}
\label{eq:deviation}
D^{i}=\frac{N_{masked}^{i}}{N_{baseline}^{i}}.
\end{equation}
$D^{i}$ denotes the deviation in the noise power for LO letter $i$, $N_{masked}^{i}$ is the noise power recorded by the homodyne detection with the mask present, and $N_{baseline}^{i}$ signifies the unmasked baseline noise power.

For the classical technique, the presence of the mask leads to varying degrees of reduction in the measured excess noise (Fig.~\ref{fig:zmaskdata}(b)).
For the quantum technique, the mask causes the measured quadrature-difference noise to increase and, for most choices of LO letter, for the squeezing to be lost.
In fact, only the correct choice of the LO letter (Z) continued to yield any squeezing.
It should be noted that the letter I in the chosen font did not produce a strong enough LO to elevate the single beam shot noise significantly above the electronic noise floor, so it is not included in Fig.~\ref{fig:zmaskdata}(b)-(c).

Although both the classical and quantum estimation techniques correctly suggest that the letter Z is the best choice of LO, the quantum technique provides a higher degree of confidence.
Uncertainties in the deviation incorporate contributions from both the baseline and masked noise measurements.
These uncertainties are plotted along with their associated changes in squeezing and excess noise in Fig.~\ref{fig:zmaskdata}.
For the quantum case (Fig.~\ref{fig:zmaskdata}(c)), the correct choice of LO is separated from the next closest choice (the letter T) by over 7 standard deviations.
For the classical case, however, the quadrature noises for the letters T and Z lie within their respective uncertainties.

\section{Conclusion}
In conclusion, we have demonstrated the ability to estimate the shape of a binary intensity mask using homodyne detection of the noise of vacuum-squeezed twin beams used to illuminate the object.
In the classical and quantum techniques described, the mask is exposed to exactly the same state of light, which contains too few photons for direct detection of the amplitude quadrature and too many photons for current photon counters.
These techniques might therefore prove useful for imaging applications where a minimal exposure to light is critical (e.g. when dealing with low optical damage thresholds or when it is desirable that the detection go unnoticed).
A promising application is related to imaging of biological samples under controlled conditions where squeezing could be maintained and the exposure of the sample minimized \cite{nasr_quantum_2009}.

The classical noise imaging technique exploits the fact that, taken individually, each of the twin beams behaves like a thermal state of a light characterized by uniform excess quadrature noise.
Although the excess noise properties of a single beam can be used, we have demonstrated that the quantum correlations of the twin beams offer an enhancement in sensitivity for the regime of high LO-mask overlap.
As explained in \cite{marino_extracting_2012}, these correlations must be quantum for an enhancement to be possible.
Additionally, we showed that a mask can be recognized among a given alphabet of choices with increased confidence using the quantum noise imaging technique.
As one can notice from Fig.~5, the resolution with which we generate the LOs is limited.
The constraint on the resolution is the limited pump power available for the 4WM process, which places a practical upper bound on the pump's available transverse size.
Increasing the pump's transverse size at a given intensity would increase the resolution of the LOs \cite{brambilla_high-sensitivity_2008} and should improve our ability to estimate the shape of the mask.

Finally, we note that one could in principle identify a phase mask separate from, or in addition to, an intensity mask with a similar technique.
The search for a phase mask would require a scan over the LO phase in only the beam that passes through the mask.  Such a search would not be possible, however, using the ``classical technique" since for a thermal beam all phases have the same noise level.

\end{document}